# An Assembler Driven Verification Methodology (ADVM)

John S. MacBeth
Verilab GmbH
Lichtenbergerstrasse 8, Garching
85748 Munich, Germany
john.macbeth@verilab.com

Dietmar Heinz
Infineon Technologies AG
St. Martin Strasse 76
81609 Munich, Germany
dietmar.heinz@infineon.com

Ken Gray
Verilab GmbH
Lichtenbergerstrasse 8, Garching
85748 Munich, Germany
ken.gray@verilab.com

## Abstract

*This paper presents an overview of an assembler driven verification methodology (ADVM) that was created and implemented for a chip card project at Infineon Technologies AG [2]. The primary advantage of this methodology is that it enables rapid porting of directed tests to new targets and derivatives, with only a minimum amount of code refactoring. As a consequence, considerable verification development time and effort was saved.*

## 1. Introduction

Presently, the methods used for verifying complex microprocessor based System-on-Chip (SoC) devices can take many forms [1]. Such verification methods vary from high-level verification languages or tools (such as Verisity's Specman tool and 'e' language or Synopsys's Vera) to running compiled test code on the device. The advantage of the latter method, running code on the device, is that the code can be run on any simulation or emulation platform. This means that the same suite of assembler tests can be used to perform functional verification of each of the following development platforms:

**Golden Reference Model** The software simulator that is supplied to the customer for software development

**HDL-RTL Simulation** The HDL design for silicon

**HDL-Gate Level Simulation** The post-synthesis HDL design for silicon

**Hardware Accelerator** The hardware emulator used by both the customer and the embedded software teams for firmware sign-off (e.g Quickturn and IKOS)

**Bondout Silicon** The software development platform with silicon performance. It is also enhanced to include extra hardware debugging capabilities

**Product Silicon** The final product silicon for the customer

Although this method has a clear advantage in terms of crossing many simulation/emulation domains, it is restricted by the short comings of directed testing.

To verify a complex device using directed testing, many tests must be written in an attempt to cover as many functional modes of operation as possible. This means that over time a large collection of directed test code will be developed and will require re-factoring (i.e. maintenance) with each change in the specification or when migrating the test code to new derivatives.

Writing test code that crosses simulation domains should not present many problems as most platforms should execute the code in the same way. If they don't then a bug or issue has been found in that particular simulation domain. A challenge emerges however, if the aim is to re-target existing test code to a new derivative and it is this challenge that the ADVM primarily overcomes. The following section describes the test environment required to accomplish this goal.

## 2. Test Environment Structure

Abstraction is the key to dealing with change. Creating an abstraction layer between the test and the physical implementation provides a layer that can absorb any changes.

If there is only one test then such an abstraction layer would be an overhead but it is a single point from which it is possible to control change when there are many tests that share common functions. With this in mind; Figure 1 shows the basic structure of a module test environment that is composed of three layers. The test layer (highlighted in light grey) contains all of the tests associated with this particular module or class of tests. Note that this structure does not exclude the tests from using other modules.

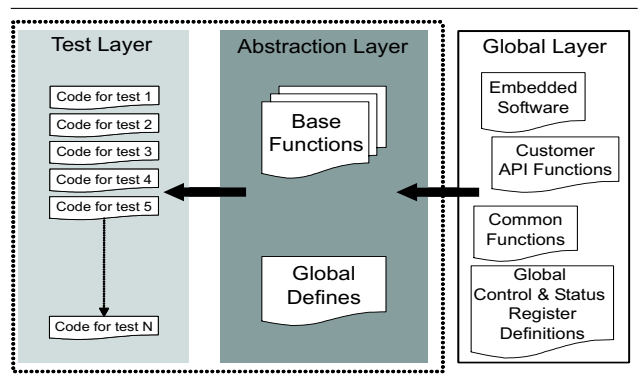

**Figure 1. Module Test Environment Structure**

The abstraction layer (highlighted in dark grey) contains two main components. The first, 'Global Defines', contains a collection of defines that are used to control the test environment. Anywhere in the test code that would have previously used a hardwired value will now be referenced in this global defines file. This file should now contain derivative specific information which can be controlled using a macro. Now the test environment can adapt automatically depending on the derivative. Similarly, the control of the test environment can be changed depending on the target simulation platform using the same technique. The second component included in the abstraction layer is a library of functions, named 'Base Functions'. Such functions are common tasks that are required by multiple tests. Once this library has been created the development time of new tests for this environment decreases considerably. A further benefit, that is not immediately obvious, is that the functions can be altered based on the derivative or the simulation target using the same technique used by the 'Global Defines' file. Critically, these functions do not contain hardwired values as they use the same 'Global Defines' file that is used by the tests. This ensures that the global defines really do control the complete test environment.

The last layer is the global layer (highlighted in white). This layer encapsulates anything that the test environment owner does not control. This might include elements such as the embedded software functions or the register definitions required for the compilers and other tools. Any functions or definitions that are located in this layer are subject to change, thus protection from change must be provided via the abstraction layer. In the case of functions, the 'Base Functions' library will wrap each of the global functions so that the tests can never call them directly. This may be thought of as overhead but consider the following scenario. A function located in the embedded software, which has been stable for months and is required by some of the current tests, has now been re-written in such a way that the input registers have been swapped around. Using an approach that does not use abstraction means that all tests that use this function need to be identified and re-factored. Alternatively, by using the ADVM, only the 'Base Functions' file needs to be re-factored, saving time and effort. The actual change can be managed by introducing extra code to swap the input parameters or by copying the original function into the 'Base Functions' and encapsulating the reference to the embedded software function.

Ultimate protection from global layer changes comes from not using any functions or definitions that exists there, but this approach is unrealistic. To deal with global layer definitions specifically, it is necessary to re-map them using the 'Global Defines' file. This protects the test environment from any global changes, e.g. name changes, to prevent the situation where a register name has been changed for a new derivative.

The test development time advantages are clear when it comes to supporting derivative change and further test development, however, there will be an initial time penalty while developing the abstraction layer. This development process is iterative and requires some re-factoring of initial tests. For example, it is not always obvious that a section of code should actually be implemented as a function and adding it later will involve some rework. Due to this iterative process between the test layer and the abstraction layer, it makes sense to release stable versions of the test environment. These stable releases can then be used by others that are required to run regressions. This is a critical point since the abstraction layer will completely control the tests and therefore cannot change during a regression.

Figure 2 shows a violation of the abstraction layer concept, by linking code directly into the tests without wrapping it in the abstraction layer. Often, it is tempting to bypass the abstraction layer, especially when under time pressure. However, by doing so, any protection from change will be lost and re-factoring of all relevant tests will be required. This situation is exactly what we are trying to avoid as this will cost the project time and needless effort.

The proposed module test environment directory structure is shown in Figure 3. Starting at the top; the test environment directory should be named after the module that is the primary focus of the test environment. Equally, the test environment can be focused on a class of tests (i.e. control and status register test) and as such the name of the test environment directory should capture this. Derivative specific names are not permitted as they will make the environment appear derivative specific.

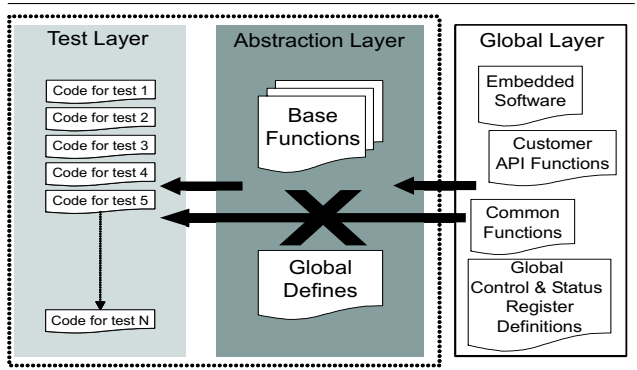

**Figure 2. Abuse of the Module Test Environment Structure**

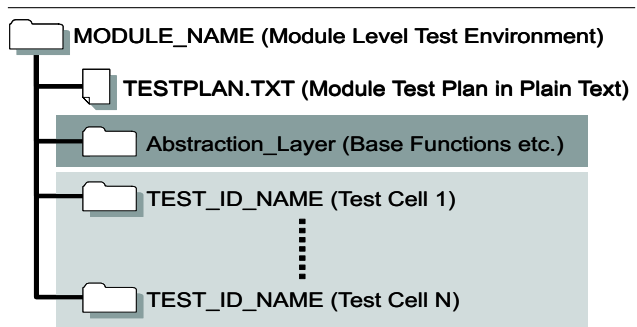

**Figure 3. Module Directory Structure**

Every test environment should contain a plain text file (illustrated as a file) that contains the test plan for the module or class of tests. The principle reason for using plain text is that it can be searched (grep'ed) easily from the command line. All of the files associated with the abstraction layer should be located in an 'Abstraction_Layer' directory (highlighted in dark grey). Such a directory will contain the library of functions that are used by all of the tests. Each test cell will contain a link in its source directory that points to the required files located in the base functions directory. The remaining directories are the test cells that have identical file directory structures so that consistency exists across the whole environment.

Looking towards the future, this test environment structure provides the ability to generate constrained-random instances of the 'Global Defines' file from a higher level language such as Specman e, Perl or even C/Cpp. Furthermore, the Base_Functions library could be considered as a library of assembler code functions that can be called or linked into some higher level language.

While this test environment structure addresses many of the issues relating to rapid porting to new derivatives, it is not the complete solution. However, in terms of the system verification environment this test environment structure represents a vital component of the complete solution.

## 3. Complete Test Environment Structure

The complete test environment structure is shown in Figure 4 and is composed of multiple module level test environments with a shared global layer.

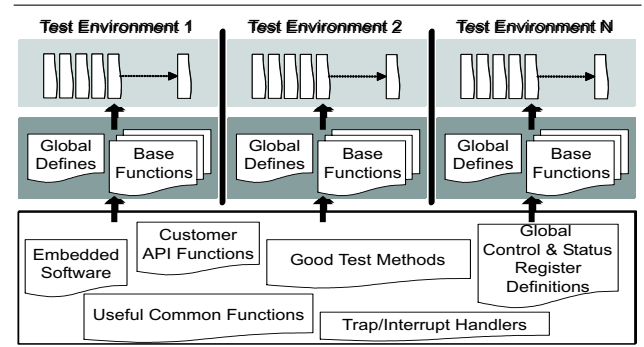

**Figure 4. The Complete Test Environment Structure**

Each test environment is isolated from any other and the only way for code to be shared is via the globals layer (highlighted in white). This is deliberate and prevents reliance on code that is not controlled by the test environment developer. This is an important point as any dependencies that cannot be fully controlled via the abstraction layer will cause problems if they are changed (sometimes without the test environment owner knowing about it). Another reason for the separation of each test environment is ownership; in particular the test environment owner is responsible for the coding, control and release of their environment. In terms of revision control, each module or test class owner will be responsible for releasing a working version of their test environment. Such releases can be controlled by revision control software in the form of a label. This label can then be given to anyone who requires it for running regressions at the test environment level. Extending this to the system regression level, it is now possible to release an instance of the complete test environment for regressions by creating a label composed of sub-labels for each environ-ment. There should be a single person responsible for the release of a complete regression environment so that they can effectively manage any issues that arise from regressions. This release mechanism is vital as any changes made to the abstraction layer will have a global effect on the tests. Thus, the test environment is not stable during any development of the abstraction layer, unless frozen via a release label.

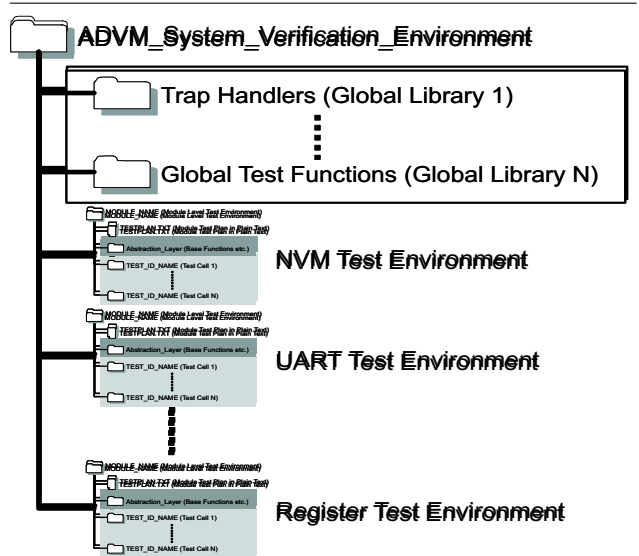

**Figure 5. System Directory Structure**

Figure 5 demonstrates how this system level environment is implemented from a file system point of view. The complete system verification environment should located in one main directory. Inside this master directory are some sub-directories that contain libraries of global functions or definitions (highlighted in white). Such libraries are normally outwith the module test owners control and as such are considered to be located in the global layer. Following on from these global library directories, are the test environment directories that were shown above in Figure 3.

The complete verification methodology relies on the abstraction layer and its ability to hide, from the test layer, any changes to the design or to the globals layer. The following section presents some code examples describing how this abstraction layer is achieved.

## 4. Abstraction Layer Code Examples

As previously highlighted, the abstraction layer is the most critical aspect of this design philosophy. Ideally, it is a collection of code that provides complete control of the test code contained in the test layer. Creating such a means of control provides the ability to change the operation of tests depending on derivative or target and it is all controlled from the abstraction layer. Therefore a single point of change is possible via the abstraction layer instead of test re-factoring or worse still, test re-creation.

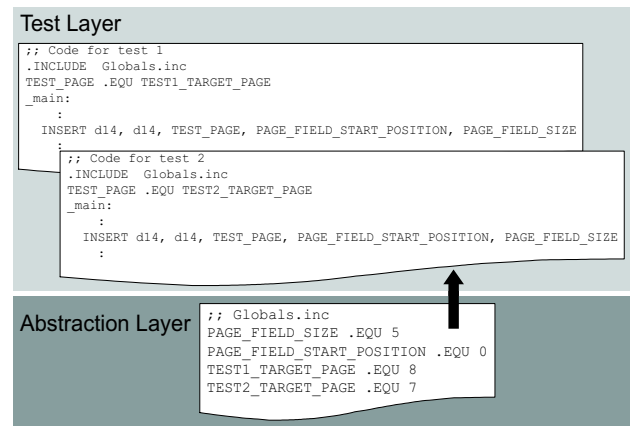

**Figure 6. Simple Code Example 1**

Figure 6 shows a simple code example that illustrates the use of the abstraction layer for test control and for protection against specification or derivative changes. In the test layer (displayed in light grey) there are two tests that create a data value (stored in data register d14) that will be written to a module control register. The creation of such a value relies on two things; the location of the control bits in the control register (defined by PAGE_FILE_SIZE and PAGE_FIELD_START_POSITION) and the value to be used for this control field (defined by TEST1_TARGET_PAGE and TEST2_TARGET_PAGE). Clearly, the code is much more readable using these well named defines but also note that the control value of the tests are located locally within each test. This offers the test writer some local control for debugging the test and provides the ability to focus the test on a specific corner case. This local control is only a place holder as the real control comes from the abstraction layer via the globals.inc file. Using this globals file it is possible to control both tests without actually changing the test code. Also included in the globals file are the two defines that deal with the location and size of the control field that is being accessed (PAGE_FILE_SIZE and PAGE_FIELD_START_POSITION). Initially these two control defines appear redundant but consider a specification change where the location of these control bits have been shifted by one. Now this change can be absorbed easily by modifying only the globals file instead of having to edit each test file (assuming that the af-fected tests are known). The next type of change that can occur is that a new derivative is produced but this version of the module is now capable of handling more pages. To handle these extra pages the page control field size has increased by one bit. This is not a problem because the globals file can be used to specify derivative specific information (allowed only in the abstraction layer) so such a change can be handled without a re-write of the tests (i.e. the PAGE_FILE_SIZE define can be changed from 5 to 6 for this derivative).

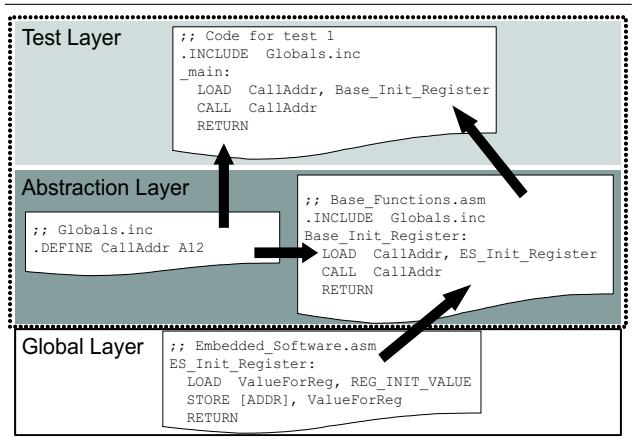

**Figure 7. Simple Code Example 2**

Another simple code example is shown in Figure 7. In this case a base functions file (base_functions.asm) is also shown in the abstraction layer and is really a library of functions required by the tests. For this particular example a function that has global scope (i.e. not under the module test environment control) is required by the test. Normally, this code would be included and called directly, but this solution would bypass the abstraction layer and would cause problems if it ever changed (remember that it is not under your control). To solve this issue the function is wrapped by the abstraction layer then this encapsulated function is called from the test. Some examples of typical changes that could occur are; the function name, the input and output parameters or the actual code itself. The abstraction layer does not stop the change taking place but provides a single point to handle it. The solution might be as simple as breaking the link to the global layer and inserting a specific version of the function or it could just be a re-map of the inputs and outputs respectively.

## 5. Conclusions

This paper provided an overview of the assembler driven verification methodology (ADVM) that was deployed as a solution and enhancement to Infineon's existing verification environment.

As with any form of software development (complex test environments are just software projects) the code must be written with a focus on the present requirements but also with an awareness of possible future needs. This requires some up-front thinking about which aspects of the test may change. Such changes may include; the derivative, the target simulation platform, the specification or the focus of the test (control for corner cases). The need to code in this way can be viewed as a disadvantage as it is not possible to get all of this correct first time. As a result the test environment will require some iterative development between the test and abstraction layers.

Clearly, this style of coding introduces some overhead and possibly some redundancy (not all of the predicted changes will happen) but with more readable and controllable code this overhead is acceptable. Consequently, there is an initial time penalty associated with the development of the test environment. However, this time is easily recovered on first reuse with a new target platform or derivative.

Any disadvantages are outweighed by the advantages that the ADVM offers, namely:

- All changes are resolved by the abstraction layer
- Rapid porting to new derivatives is achieved since the abstraction layer is inherited by all tests
- Provides a consistent method for the creation of tests that enables easier training of new personnel and easier debugging of tests written by another author
- Provides a better method of test control that allows specific corner cases to be investigated (as the need arises)
- Once the base functions for each environment have been created the test development time is significantly reduced
- The existing test environment is not lost, but can be replaced gradually
- Provides the possibility of generating more complex test scenarios using high level languages

In conclusion, the application of the ADVM has greatly improved our existing and future verification efforts.

## 6. Acknowledgements

The authors acknowledge the support given to this work by the SLE88 verification team, namely: Steffen Heinkel, Hans-Hubert Kallus, Josef Huber, Wolfgang Gaertner and Ulrike Pfannkuchen.